\documentclass{amsart}
\usepackage{amsfonts}
\usepackage{amsmath}
\usepackage{amssymb}

\theoremstyle{plain}
\newtheorem{theorem}{Theorem}[section]

\newtheorem{proposition}[theorem]{Proposition}
\theoremstyle{definition}

\theoremstyle{remark}
\newtheorem{remark}[theorem]{Remark}
\numberwithin{equation}{section}

\input{tcilatex}

\begin{document}
\title[The relativistic electrodynamics least action principles]{The
relativistic electrodynamics least action principles revisited: new
charged point particle and hadronic string models analysis}
\author{N.N. Bogolubov (jr.)}
\address{V.A. Steklov Mathematical Institute of RAS, Moscow, Russian
Federation\\
and\\
The Abdus Salam International Centre of Theoretical Physics, Trieste, Italy}
\email{nikolai\_bogolubov@hotmail.com}
\author{A.K. Prykarpatsky}
\address{The AGH University of Science and Technology, Department of Applied
Mathematics, Krakow 30059 Poland\\
and\\
Ivan Franko Pedagogical State University, Drohobych, Lviv region, Ukraine \ }
\email{pryk.anat@ua.fm, prykanat@cybergal}
\author{U. Taneri}
\address{Department of Applied Mathematics and Computer Science, Eastern
Mediterranean University EMU, Famagusta, North Cyprus\\
and\\
Kyrenia American University GAU, Institute of Graduate Studies, Kyrenia,
North Cyprus }
\email{ufuk.taneri@gmail.com}
\date{present}

\begin{abstract}
The classical relativistic least action principle is revisited from the
vacuum field theory approach. New physically motivated versions of
relativistic Lorentz type forces are derived, a new relativistic hadronic
string model is proposed and analyzed in detail.
\end{abstract}

\maketitle

\section{Introduction}

\subsection{The classical relativistic electrodynamics backgrounds: a
charged point particle analysis}

It is well known \cite{LL,Fe1,Pa,BN} that the relativistic least action
principle for a point charged particle $q$ in the Minkovski space-time $%
M^{4}\simeq \mathbb{E}^{3}\times \mathbb{R}$ can be formulated on a time
interval $[t_{1},t_{2}]\subset \mathbb{R}$ (in the light speed units) as
\begin{eqnarray}
\delta S^{(t)} &=&0,\text{ \ \ }S^{(t)}:=\int_{\tau (t_{1})}^{\tau
(t_{2})}(-m_{0}d\tau -q<\mathcal{A},dx>_{M^{4}})=  \notag \\
&=&\int_{s(t_{1})}^{s(t_{2})}(-m_{0}<\dot{x},\dot{x}>_{M^{4}}^{1/2}-q<%
\mathcal{A},\dot{x}>_{M^{4}})ds.  \label{A1.1}
\end{eqnarray}%
Here $\delta x(s(t_{1}))=0=\delta x(s(t_{2}))$ are the boundary constraints,
$m_{0}\in \mathbb{R}_{+}$ is the so called particle rest mass, the 4-vector $%
x:=(r,t)\in M^{4}$ is the particle location in $M^{4},$ $\dot{x}:=dx/ds\in
T(M^{4})$ is the particle 4-vector velocity with respect to a laboratory
reference system $\mathcal{K},$ parameterized by means of the Minkovski
space-time parameters $(r,s(t))\in M^{4}$ and related to each other by means
of the infinitesimal Lorentz interval relationship
\begin{equation}
d\tau :=<dx,dx>_{M^{4}}^{1/2}:=ds<\dot{x},\dot{x}>_{M^{4}}^{1/2},
\label{A1.1a}
\end{equation}%
$\mathcal{A}\in $ $T^{\ast }(M^{4})$ is an external electromagnetic 4-vector
potential, satisfying the classical Maxwell equations \cite{Pa,LL,Fe1}. the
sign $<\cdot ,\cdot >_{\mathcal{H}}$ \ means, in general, the corresponding
scalar product in a finite-dimensional vector space $\mathcal{H}$ and $%
T(M^{4}),T^{\ast }(M^{4})$ \ are, respectively, the tangent and cotangent
spaces \cite{AM,Ar,Th,DNF,HPP} to the Minkovski space $M^{4}.$ In
particular, $<x,x>_{M^{4}}:=t^{2}-<r,r>_{\mathbb{E}^{3}}$ for any $%
x:=(r,t)\in M^{4}.$

The subintegral expression in (\ref{A1.1})
\begin{equation}
\mathcal{L}^{(t)}:=-m_{0}<\dot{x},\dot{x}>_{M^{4}}^{1/2}-q<\mathcal{A},\dot{x%
}>_{M^{4}}  \label{A1.2}
\end{equation}%
is the related Lagrangian function, whose first part is proportional to the
particle world line length with respect to the proper rest reference system $%
\mathcal{K}_{r}$ and the second part is proportional to the pure
electromagnetic particle-field interaction with respect to the Minkovski
laboratory reference system $\mathcal{K}.$ Moreover, the positive rest mass
parameter \ $m_{0}\in \mathbb{R}_{+}$ is introduced into (\ref{A1.2}) as an
external physical ingredient, also describing the point particle\ \ with
respect to the proper rest reference system $\mathcal{K}_{r}.$ \ The
electromagnetic 4-vector potential $\mathcal{A}\in T^{\ast }(M^{4})$ $\ $is
at the same time expressed as a 4-vector, constructed and measured with
respect to the Minkovski laboratory reference system $\mathcal{K}$ that
looks from physical point of view enough controversial, since the action
functional (\ref{A1.1}) is forced to be extremal with respect to the
laboratory reference system $\mathcal{K}.$ This, in particular, means that
the real physical motion of our charged point particle, being realized with
respect to the proper rest reference system $\mathcal{K}_{r},$ somehow
depends on an external observation data \cite{Fe2,Fa,Lo,Lo1,Lo2,Lo3,Br} with
respect to the occasionally chosen laboratory reference system $\mathcal{K}.$
\ This aspect was never discussed in the physical literature except of very
interesting reasonings by R. Feynman in \cite{Fe1}, who argued that the
relativistic expression for the classical Lorentz force has a physical sense
only with respect to \ the Euclidean rest \ reference system $\mathcal{K}%
_{r} $ variables $(r,\tau )\in \mathbb{E}^{4}$ related with the Minkovski
laboratory reference system $\mathcal{K}$ parameters $(r,t)\in M^{4}$ by
means of the infinitesimal relationship
\begin{equation}
d\tau :=<dx,dx>_{M^{4}}^{1/2}=dt(1-u^{2})^{1/2},  \label{A1.2a}
\end{equation}%
where $u:=dr/dt\in T(\mathbb{E}^{3})$ is the point particle velocity with
respect to the reference system $\mathcal{K}.$

It is worth to point out here that to be correct, it would be necessary to
include still into the action functional the additional part describing the
electromagnetic field itself. But this part is not taken into account, since
there is generally assumed \cite{Kl,Kle,B-W,Wi,B-B,We,Me,Me1,Gr,Ne} that the
charged particle influence on the electromagnetic field is negligible. This
is true, if the particle charge value $\ q$ is very small but the support $%
supp\mathcal{A\subset }$ $M^{4}$ of the electromagnetic 4-vector potential
is compact. It is clear that in the case of two interacting to each other
charged particles the above assumption can not be applied, as it is
necessary to take into account the relative motion of two particles and the
respectively changing interaction energy. This aspect of \ the action
functional choice problem appears to be very important when one tries to
analyze the related Lorentz type \ forces exerted by charged particles on
each other. We will return to this problem in a separate section below.

Having calculated the least action condition (\ref{A1.1}), we easily obtain
from (\ref{A1.2}) the classical relativistic dynamical equations
\begin{eqnarray}
dP/ds &:&=-\partial \mathcal{L}^{(t)}/\partial x=-q\nabla _{x}<\mathcal{A},%
\dot{x}>_{M^{4}},  \label{A1.3} \\
P &:&=-\partial \mathcal{L}^{(t)}/\partial \dot{x}=m_{0}\dot{x}<\dot{x},\dot{%
x}>_{M^{4}}^{-1/2}+q\mathcal{A},  \notag
\end{eqnarray}%
where by $P\in T^{\ast }(M^{4})$ we denoted the common particle-field
momentum of the interacting system.

Now at $s=t\in \mathbb{R}$ by means of the standard infinitesimal change of
variables (\ref{A1.2a}) \ we can easily obtain from (\ref{A1.3}) the
classical Lorentz force expression%
\begin{equation}
dp/dt=qE+qu\times B  \label{A1.6}
\end{equation}%
with the relativistic particle momentum and mass
\begin{equation}
p:=mu,\text{ \ }\ m:=m_{0}(1-u^{2})^{-1/2},  \label{A1.7}
\end{equation}%
respectively, the electric field
\begin{equation}
E:=-\partial A/\partial t-\nabla \varphi  \label{A1.8}
\end{equation}%
and the magnetic field
\begin{equation}
B:=\nabla \times A,  \label{A1.9}
\end{equation}%
where we have expressed the electromagnetic 4-vector potential as $\mathcal{A%
}:=(A,\varphi )\in T^{\ast }(M^{4}).$

The Lorentz force (\ref{A1.6}), owing to our preceding assumption, means the
force exerted by the external electromagnetic field on our charged point
particle, whose charge $q$ is so negligible that it does not exert the
influence on the field. This fact becomes very important if we try to make
use of the Lorentz force expression (\ref{A1.6}) for the case of two
interacting to each other charged particles, since then one can not assume
that our charge $q$ exerts negligible influence on other charged particle.
Thus, the corresponding Lorentz force between two charged particles should
be suitably altered. Nonetheless, the modern physics \cite%
{BS,BS1,Di,LL,BjD,BD,DJ,JP,BPZZ,Ba,Ja} did not make this naturally
needed \ Lorentz force modification and there is everywhere used the
classical expression (\ref{A1.6}). This situation was observed and
analyzed concerning the related physical aspects in \cite{Re},
having shown that the electromagnetic Lorentz force between two
moving charged particles can be modified in such a way that it
ceases to be dependent on their relative motion contrary to the
classical relativistic case.

To the regret, the least action principle approach to analyzing the Lorentz
force structure was in \cite{Re} completely ignored that gave rise to some
incorrect and physically not motivated statements concerning mathematical
physics backgrounds of the modern electrodynamics. To make the problem more
transparent we will analyze it in the section below from the vacuum field
theory approach recently devised in \cite{BPT,BPT1,BP}.

\subsection{The least action principle analysis}

Consider the least action principle (\ref{A1.1}) and observe that the
extremality condition
\begin{equation}
\delta S^{(t)}=0,\text{ \ \ \ \ }\delta x(s(t_{1}))=0=\delta x(s(t_{2})),
\label{A2.1}
\end{equation}%
is calculated with respect to the laboratory reference system $\mathcal{K},$
whose point particle coordinates $(r,t)\in M^{4}$ are parameterized by means
\ of an arbitrary parameter $s\in \mathbb{R}$ owing to expression (\ref%
{A1.1a}). Recalling now the definition of the invariant proper rest
reference system $\mathcal{K}_{r}$ time parameter (\ref{A1.2a}), we obtain
that at the critical parameter value $s=\tau \in \mathbb{R}$ the action
functional (\ref{A1.1}) on the fixed interval $[\tau _{1},\tau _{2}]\subset
\mathbb{R}$ turns into
\begin{equation}
S^{(t)}=\int\limits_{\tau _{1}}^{\tau _{2}}(-m_{0}-q<\mathcal{A},\dot{x}%
>_{M^{4}})d\tau  \label{A2.2}
\end{equation}%
under the additional constraint
\begin{equation}
<\dot{x},\dot{x}>_{M^{4}}^{1/2}=1,  \label{A2.2a}
\end{equation}%
where, by definition, $\ \dot{x}:=dx/d\tau ,$ $\tau \in \mathbb{R}.$

The expressions (\ref{A2.2}) and (\ref{A2.2a}) need some comments since the
corresponding to (\ref{A2.2}) Lagrangian function
\begin{equation}
\mathcal{L}^{(t)}:=-m_{0}-q<\mathcal{A},\dot{x}>_{M^{4}}  \label{A2.3}
\end{equation}%
depends only virtually on the unobservable rest mass parameter $m_{0}\in
\mathbb{R}$ and, evidently, it has no \ direct impact into the resulting
particle dynamical equations following from the condition $\delta S^{(t)}=0.$
\ Nonetheless, the rest mass springs up as a suitable Lagrangian multiplier
owing to the imposed constraint (\ref{A2.2a}). To demonstrate this consider
the extended Lagrangian function (\ref{A2.3}) in the form
\begin{equation}
\mathcal{L}_{\lambda }^{(t)}:=-m_{0}-q<\mathcal{A},\dot{x}>_{M^{4}}-\lambda
(<\dot{x},\dot{x}>_{M^{4}}^{1/2}-1),  \label{A2.3.1}
\end{equation}%
where $\lambda \in \mathbb{R}$ is a suitable Lagrangian multiplier. The
resulting Euler equations look as%
\begin{eqnarray}
P_{r} &:&=\partial \mathcal{L}_{\lambda }^{(t)}/\partial \dot{r}=qA+\lambda
\dot{r},\text{ \ }P_{t}:=\partial \mathcal{L}_{\lambda }^{(t)}/\partial \dot{%
t}=-q\varphi -\lambda \dot{t},\text{ }  \notag \\
\text{\ }\partial \mathcal{L}_{\lambda }^{(t)}/\partial \lambda &=&<\dot{x},%
\dot{x}>_{M^{4}}^{1/2}-1=0,\text{ \ }dP_{r}/d\tau =q\nabla _{r}<A,\dot{r}>_{%
\mathbb{E}^{3}}-q\dot{t}\nabla _{r}\varphi ,  \notag \\
\text{ \ }dP_{t}/d\tau &=&q<\partial A/\partial t,\dot{r}>_{\mathbb{E}^{3}}-q%
\dot{t}\partial \varphi /\partial t,  \label{A2.3.2}
\end{eqnarray}%
giving rise, owing to relationship (\ref{A1.2a}), to the following dynamical
equations:

\begin{equation}
\frac{d}{dt}(\lambda u\dot{t})=qE+qu\times B,\text{ \ }\frac{d}{dt}(\lambda
\dot{t})=q<E,u>_{\mathbb{E}^{3}},  \label{A2.3.3}
\end{equation}%
where we denoted by
\begin{equation}
E:=-\partial A/\partial t-\nabla \varphi ,\text{ \ }B=\nabla \times A
\label{A2.3.4}
\end{equation}%
the corresponding electric and magnetic fields. As a simple consequence of (%
\ref{A2.3.3}) one obtains
\begin{equation}
\frac{d}{dt}\ln (\lambda \dot{t})+\frac{d}{dt}\ln (1-u^{2})^{1/2}=0,
\label{A2.3.5}
\end{equation}%
being equivalent for all $t\in \mathbb{R},$ owing to relationship (\ref%
{A1.2a}), \bigskip to the relationship
\begin{equation}
\lambda \dot{t}(1-u^{2})^{1/2}=\lambda :=m_{0,}  \label{A2.3.6}
\end{equation}%
where $m_{0}\in \mathbb{R}_{+}$ is a constant, which could be interpreted as
the rest mass of our charged point particle $q.$ Really, the first equation
of (\ref{A2.3.3}) can be rewritten as
\begin{equation}
dp/dt=qE+qu\times B,  \label{A2.3.7}
\end{equation}%
where we denoted
\begin{equation}
p:=mu,\text{ }m:=\lambda \dot{t}=m_{0}(1-u^{2})^{-1/2},  \label{A2.3.8}
\end{equation}%
coinciding exactly with that of (\ref{A1.2a}).

Thereby, we retrieved here all of the results obtained in section
above, making use of the action functional (\ref{A2.2}), represented
with
respect to the rest reference system $\mathcal{K}_{r}$ under constraint \ (%
\ref{A2.2a}). \ During these derivations, we faced with a very delicate
inconsistency property of definition \ of the action functional $S^{(t)},$
defined with respect to the rest reference system $\mathcal{K}_{r},$ but
depending on the external electromagnetic potential function $\mathcal{A}%
:M^{4}\rightarrow T^{\ast }(M^{4}),$ constructed exceptionally with respect
to the laboratory reference system $\mathcal{K}.$ Namely, this potential
function, as a physical observable quantity, is defined and, \ \
respectively, measurable only with respect to the fixed laboratory reference
system $\mathcal{K}.$ This, in particular, means that a physically
reasonable action functional should be constructed by means of an expression
strongly calculated within the laboratory reference system $\mathcal{K}$ by
means of coordinates $(r,t)\in M^{4}$ and later suitably transformed subject
to the rest reference system $\mathcal{K}_{r}$ coordinates $(r,\tau )\in
\mathbb{E}^{4},$ respective for the real charged point particle $q$ motion.
Thus, the corresponding action functional, in reality, should be from the
very beginning written as
\begin{equation}
S^{(\tau )}=\int\limits_{t(\tau _{1})}^{t(\tau _{2})}(-q<\mathcal{A},\dot{x}%
>_{\mathbb{E}^{3}})dt,  \label{A2.4}
\end{equation}%
where $\dot{x}:=dx/dt,$ $t\in \mathbb{R},$ being calculated on some time
interval $[t(\tau _{1}),t(\tau _{2})]\subset \mathbb{R},$ suitably related
with the proper motion of the charged point particle $q$ on the true time
interval $[\tau _{1},\tau _{2}]\subset \mathbb{R}$ with respect to the rest
reference system $\mathcal{K}_{r}$ and whose charge value is assumed so
negligible that it exerts no influence on the external electromagnetic
field. The problem now arises: how to compute correctly the variation $%
\delta S^{(\tau )}=0$ of the action functional (\ref{A2.4})?

To reply to this question we will turn to the Feynman reasonings from \cite%
{Fe1}, where he argued, when deriving the relativistic Lorentz force
expression, that the real charged particle dynamics can be
physically not ambiguously determined only with respect to the rest
reference system time parameter. Namely, \ Feynman wrote: "...we
calculate a growth $\Delta x$ for a small time interval $\Delta t.$
But in the other reference system the
interval $\Delta t$ may correspond to changing both $t^{\prime }$ and $%
x^{\prime },$ thereby at the change of the only $t^{\prime }$ the suitable
change of $x$ will be\ other... Making use of the quantity $d\tau $ one can
determine a good differential operator \ $d/d\tau ,$ as it is invariant with
respect to the Lorentz reference systems transformations". This means that
if our charged particle $q$ moves in the Minkovski space $M^{4}$ during the
time interval $[t_{1},t_{2}]\subset \mathbb{R}$ with respect to the
laboratory reference system $\mathcal{K},$ its proper real and invariant
time of motion with respect to the rest reference system $\mathcal{K}_{r}$
will be respectively $[\tau _{1},\tau _{2}]\subset \mathbb{R}.$

As a corollary of the Feynman reasonings, we arrive at the necessity to
rewrite the action functional (\ref{A2.4}) as%
\begin{equation}
S^{(\tau )}=\int\limits_{\tau _{1}}^{\tau _{2}}(-q<\mathcal{A},\dot{x}%
>_{M^{4}})d\tau ,\text{ \ }\delta x(\tau _{1})=0=\delta x(\tau _{2}),
\label{A2.5}
\end{equation}%
where $\dot{x}:=dx/d\tau ,$ $\tau \in \mathbb{R},$ under the additional
constraint
\begin{equation}
<\dot{x},\dot{x}>_{M^{4}}^{1/2}=1,  \label{A2.5a}
\end{equation}%
being equivalent to the infinitesimal transformation (\ref{A1.2a}).
Simultaneously the proper time interval $[\tau _{1},\tau _{2}]\subset
\mathbb{R}$ \ is mapped on the time interval $[t_{1},t_{2}]\subset \mathbb{R}
$ by means of the infinitesimal transformation
\begin{equation}
dt=d\tau (1+\dot{r}^{2})^{1/2},  \label{A2.6}
\end{equation}%
where $\dot{r}:=dr/d\tau ,$ $\tau \in $\ $\mathbb{R}.$ Thus, we can now pose
the true least action problem equivalent to (\ref{A2.5}) as
\begin{equation}
\delta S^{(\tau )}=0,\text{ \ \ \ \ \ }\delta r(\tau _{1})=0=\delta r(\tau
_{2}),  \label{A2.6a}
\end{equation}%
where the functional
\begin{equation}
S^{(\tau )}=\int\limits_{\tau _{1}}^{\tau _{2}}[-\bar{W}(1+\dot{r}%
^{2})^{1/2}+q<A,\dot{r}>_{\mathbb{E}^{3}}]d\tau  \label{A2.7}
\end{equation}%
is characterized by the Lagrangian function
\begin{equation}
\mathcal{L}^{(\tau )}:=-\bar{W}(1+\dot{r}^{2})^{1/2}+q<A,\dot{r}>_{\mathbb{E}%
^{3}}.  \label{A2.8}
\end{equation}%
Here we denoted, for further convenience, $\bar{W}:=q\varphi ,$ \ being the
suitable vacuum field \cite{BPT,BPT1,BP,Re} potential function. The
resulting Euler equation gives rise to the following relationships%
\begin{eqnarray}
P &:&=\partial \mathcal{L}^{(\tau )}/\partial \dot{r}=-\bar{W}\dot{r}(1+\dot{%
r}^{2})^{-1/2}+qA,  \label{A2.9} \\
dP/d\tau &:&=\partial \mathcal{L}^{(\tau )}/\partial r=-\nabla \bar{W}(1+%
\dot{r}^{2})^{1/2}+q\nabla <A,\dot{r}>_{\mathbb{E}^{3}}.  \notag
\end{eqnarray}%
Making now use once more of the infinitesimal transformation (\ref{A2.6})
and the crucial dynamical particle mass definition \cite{BPT,BP,Re} (in the
light speed units )
\begin{equation}
m:=-\bar{W},  \label{A2.9a}
\end{equation}%
we can easily rewrite equations (\ref{A2.9}) with respect to the parameter $%
t\in \mathbb{R}$ as the \ classical relativistic Lorentz force:%
\begin{equation}
dp/dt=qE+qu\times B,  \label{A2.10}
\end{equation}%
where we denoted
\begin{eqnarray}
p &:&=mu,\text{ \ \ \ \ \ \ }u:=dr/dt,\text{ }  \label{A2.11} \\
B &:&=\nabla \times A,\text{ \ \ }E:=-q^{-1}\nabla \bar{W}-\partial
A/\partial t.  \notag
\end{eqnarray}%
Thus, we obtained once more the relativistic Lorentz force expression (\ref%
{A2.10}), but slightly different from (\ref{A1.6}), since the classical
relativistic momentum $\ $expression of (\ref{A1.7}) \ does not completely
coincide with our modified relativistic momentum expression
\begin{equation}
p=-\bar{W}u,  \label{A2.12}
\end{equation}%
depending strongly on the scalar vacuum field potential function $\bar{W}%
:M^{4}\rightarrow \mathbb{R}.$ But if to recall here that our action
functional (\ref{A2.5}) was written under the assumption that \ the particle
charge value $q$ is negligible and \ not exerting the essential influence on
the electromagnetic field source, we can make use of the before obtained in
\cite{BP,BPT,Re} result, that the vacuum field potential function $\bar{W}%
:M^{4}\rightarrow \mathbb{R},$ owing to (\ref{A2.10})-(\ref{A2.12}),
satisfies as $q\rightarrow 0$ the dynamical equation
\begin{equation}
d(-\bar{W}u)/dt\simeq -\nabla \bar{W},  \label{A2.13}
\end{equation}%
whose solution will be exactly the expression%
\begin{equation}
-\bar{W}=m_{0}(1-u^{2})^{-1/2},\text{ }m_{0}=-\left. \bar{W}\right\vert
_{u=0}.  \label{A2.14}
\end{equation}%
Thereby, we have arrived, owing to (\ref{A2.14}) and (\ref{A2.12}), at the
almost full coincidence of our result (\ref{A2.10}) for the relativistic \
Lorentz force with that of (\ref{A1.6}) under the condition $q\rightarrow 0.$

The obtained above results and inferences we will formulate as the following
proposition.

\begin{proposition}
\label{Pr_A1.2} Under the assumption of the negligible influence of
a charged point particle $q$ on an external electromagnetic field
source a
true physically reasonable action functional can be given by expression (\ref%
{A2.4}), being equivalently defined with respect to the rest reference
system $\mathcal{K}_{r}$ in form (\ref{A2.5}),(\ref{A2.5a}). The resulting
relativistic Lorentz force (\ref{A2.10}) coincides almost exactly with that
of (\ref{A1.6}), obtained from the classical Einstein type action functional
(\ref{A1.1}), but the momentum expression (\ref{A2.12}) differs from the
classical expression (\ref{A1.7}), taking into account the related vacuum
field potential interaction energy impact.
\end{proposition}

\bigskip As an important corollary we make the following.

\begin{corollary}
\label{Cor_A2.2}The Lorentz force expression (\ref{A2.10}) \ should be in due course corrected in the case when the weak charge $q$ influence
assumption made above does not hold.
\end{corollary}

\begin{remark}
\label{Rem_A2.3}Concerning the infinitesimal relationship (\ref{A2.6}) one
can observe that it reflects the Euclidean nature of transformations $%
\mathbb{R\ni }$ $t\rightleftharpoons \tau \in $ \bigskip $\mathbb{R}.$
\end{remark}

In spite of the results obtained above by means of two different
least action principles (\ref{A1.1}) and (\ref{A2.5}), we must claim
here that the first one possesses some logical controversies, which
may give rise to unpredictable, unexplainable and even nonphysical
effects. Amongst these
controversies we mention: $i)$ the definition of Lagrangian function (\ref%
{A1.2}) as an expression, depending on the external and undefined rest mass
parameter with respect to the rest \ reference system $\mathcal{K}_{r}$ time
parameter $\tau \in \mathbb{R},$ but serving as an variational integrand
with respect to the laboratory \ reference system $\mathcal{K}$ time
parameter $t\in \mathbb{R};$ $ii)$ the least action condition (\ref{A1.1})
is calculated with respect to the fixed boundary conditions at the ends of a
time interval $[t_{1},t_{2}]\subset \mathbb{R},$ thereby the resulting
dynamics becomes strongly \ dependent on the chosen laboratory reference
system $\mathcal{K},$ what is, following the Feynman arguments \cite{Fe1,Fe2}%
, physically unreasonable; $iii)$ the resulting relativistic particle mass
and its energy depend only on the particle velocity in the laboratory
reference system $\mathcal{K},$ not taking into account the present vacuum
field potential energy, exerting not trivial action on the particle motion; $%
iv)$ the assumption concerning the negligible influence of a charged point
particle on the external electromagnetic field source is also physically
inconsistent.

\section{\protect\bigskip The charged point particle least action principle
revisited: the vacuum field theory approach}

\subsection{A free charged point particle in the vacuum medium}

\bigskip We start now from the following action functional for a charged
point particle $q$ moving with velocity $u:=dr/dt\in \mathbb{E}^{3}$ with
respect to a laboratory reference system $\mathcal{K}:$%
\begin{equation}
S^{(\tau )}:=-\int_{t(\tau _{2})}^{t(\tau _{1})}\bar{W}dt,\text{ \ \ \ \ \ }
\label{A3.1}
\end{equation}%
being defined on the time interval $[t(\tau _{1}),t(\tau _{2})]\subset
\mathbb{R}$ by means of a vacuum field potential function $\bar{W}%
:M^{4}\rightarrow \mathbb{R},$ characterizing the intrinsic properties of
the vacuum medium and its interaction with a charged point particle $q,$ \
jointly with the constraint
\begin{equation}
<\dot{\xi},\dot{\xi}>_{\mathbb{E}^{4}}^{1/2}=1,\text{ \ }  \label{A3.1a}
\end{equation}%
where $\xi :=(r,\tau )\in \mathbb{E}^{4}$ is a charged point particle
position 4-vector with respect to the proper rest reference system $\mathcal{%
K}_{r},$ $\dot{\xi}:=d\xi /dt,$ $t\in \mathbb{R}.$ As the real dynamics of
our charged point particle $q$ depends strongly only on the time interval $%
[\tau _{1},\tau _{2}]\subset \mathbb{R}$ of its own motion subject to the
rest reference system $\mathcal{K}_{r},$ we need to calculate the
extremality condition
\begin{equation}
\delta S^{(\tau )}=0,\text{ \ \ \ }\delta r(\tau _{1})=0=\delta r(\tau _{2}).
\label{A3.2}
\end{equation}%
As action functional (\ref{A3.1}) is equivalent, owing to (\ref{A3.1a}) or (%
\ref{A2.6}), to the following:%
\begin{equation}
S^{(\tau )}:=-\int_{\tau _{2}}^{\tau _{1}}\bar{W}(1+\dot{r}^{2})^{1/2}d\tau ,
\label{A3.3}
\end{equation}%
where, by definition, $\dot{r}:=dr/d\tau ,$ $\tau \in \mathbb{R},$ from (\ref%
{A3.3}) and (\ref{A3.2}) one easily obtains that
\begin{equation}
p:=-\bar{W}\dot{r}(1+\dot{r}^{2})^{-1/2},\text{ }dp/d\tau =-\nabla \bar{W}(1+%
\dot{r}^{2})^{1/2}.  \label{A3.4}
\end{equation}%
Taking into account once more \ relationship (\ref{A2.6}) we can rewrite (%
\ref{A3.4}) equivalently as
\begin{equation}
dp/dt=-\nabla \bar{W},\text{\ \ \ \ }p:=-\bar{W}u.  \label{A3.5}
\end{equation}%
If to recall the dynamic mass definition (\ref{A2.9a}), equation (\ref{A3.5}%
) turns into the Newton dynamical expression
\begin{equation}
dp/dt=-\nabla \bar{W},\text{ }p=mu.  \label{A3.6}
\end{equation}%
Having observed now that equation (\ref{A3.6}) is completely equivalent to
equation (\ref{A2.13}), we obtain right away from (\ref{A2.14}) that the
particle mass
\begin{equation}
m=m_{0}(1-u^{2})^{-1/2},\text{ }  \label{A3.7}
\end{equation}%
where
\begin{equation}
m_{0}:=-\left. \bar{W}\right\vert _{u=0}  \label{A3.7a}
\end{equation}%
is the so called particle rest mass. Moreover, since the corresponding to (%
\ref{A3.3}) Lagrangian function
\begin{equation}
\mathcal{L}^{(\tau )}:=-\bar{W}(1+\dot{r}^{2})^{1/2}  \label{A3.8}
\end{equation}%
is not degenerate, we can easily construct \cite{AM,Ar,DNF,HPP,BP} the
related conservative Hamiltonian function
\begin{equation}
\mathcal{H}^{(\tau )}=-(\bar{W}^{2}-p^{2})^{1/2},  \label{A3.9}
\end{equation}%
satisfying the canonical Hamiltonian equations%
\begin{equation}
dr/d\tau =\partial \mathcal{H}^{(\tau )}/\partial p,\text{ }dp/d\tau
=-\partial \mathcal{H}^{(\tau )}/\partial r  \label{A3.10}
\end{equation}%
and conservation conditions
\begin{equation}
d\mathcal{H}^{(\tau )}/dt=0=d\mathcal{H}^{(\tau )}/d\tau  \label{A3.11}
\end{equation}%
for all $\tau ,t\in \mathbb{R}.$ Thereby, the quantity
\begin{equation}
\mathcal{E}:=(\bar{W}^{2}-p^{2})^{1/2}  \label{A3.12}
\end{equation}%
can be naturally interpreted as the point particle total energy.

It is important to note here that energy expression (\ref{A3.12}) takes into
account both kinetic and potential energies, but the particle dynamic mass (%
\ref{A3.7}) depends only on its velocity, reflecting its free motion in \
vacuum. Moreover, since the vacuum potential function $\bar{W}%
:M^{4}\rightarrow \mathbb{R}$ is not, in general, constant, we claim that
the motion of our particle $q$ with respect to the laboratory reference
system $\mathcal{K}$ is not, in general, linear and with not constant
velocity, the situation, being already discussed before by R. Feynman in
\cite{Fe2}. Thus, we obtained the classical relativistic mass dependence on
the freely moving particle velocity (\ref{A3.7}), taking into account both
the nonconstant vacuum potential function $\bar{W}:M^{4}\rightarrow \mathbb{R%
}$ and the particle velocity $u\in \mathbb{E}^{3}.$

Mention also here that the vacuum potential function $\bar{W}%
:M^{4}\rightarrow \mathbb{R}$ itself should be simultaneously found
by means of a suitable solution to the Maxwell equation $\partial
^{2}W/\partial t^{2}-\Delta W=\rho ,$ where $\rho \in \mathbb{R}$ is
an ambient charge density and, by definition, $\ \bar{W}(r(t))$ $:=$
$\lim_{r\rightarrow r(t)}\left. W(r,t)\right\vert ,$ with $r(t)\in
\mathbb{E}^{3}$ being the position of the charged point particle at
a time moment $t\in \mathbb{R}.$

Return now to expression (\ref{A3.1}) and rewrite it in the following
invariant form%
\begin{equation}
S^{(\tau )}=-\int_{s(\tau _{1})}^{s(\tau _{2})}\bar{W}<\dot{\xi},\dot{\xi}>_{%
\mathbb{E}^{4}}^{1/2}ds,  \label{A3.13}
\end{equation}%
where, by definition, $s\in \mathbb{R}$ parameterizes the particle world
line related with the laboratory reference system $\mathcal{K}$ time
parameter $t\in \mathbb{R}$ by means of the Euclidean infinitesimal
relationship
\begin{equation}
dt:=<\dot{\xi},\dot{\xi}>_{\mathbb{E}^{4}}^{1/2}ds.  \label{A3.14}
\end{equation}%
It is easy to observe that at $s=t\in \mathbb{R}$ functional (\ref{A3.13})
turns into (\ref{A3.1}) and (\ref{A3.1a}). The action functional (\ref{A3.13}%
) is to be supplemented with the boundary conditions
\begin{equation}
\delta \xi (s(\tau _{1}))=0=\delta \xi (s(\tau _{2})),  \label{A3.15}
\end{equation}%
which are, obviously, completely equivalent to those of (\ref{A3.2}), since
the mapping $\mathbb{R}$ $\ni s\rightleftharpoons t\in \mathbb{R},$ owing to
definition (\ref{A3.14}) is one-to-one.

Having calculated the least \ action condition $\delta S^{(\tau )}=0$ \
under constraints (\ref{A3.15}), one easily obtains the same equation (\ref%
{A3.5}) and relationships (\ref{A3.7}), (\ref{A3.12}) for the particle
dynamical mass and its conservative energy, respectively.

\subsection{The charged point particle electrodynamics}

We would like to generalize the results obtained above for a free point
particle in the vacuum medium on the case of a charged point particle $q$
interacting with external charged point particle $q_{f},$ both moving with
respect to a laboratory reference system $\mathcal{K}.$ Within the vacuum
field theory approach, devised in \cite{BPT,BPT1,BP}, it is naturally to
reduce the formulated problem to that considered above, having introduced
the reference system $\mathcal{K}_{f}$ moving with respect to the reference
system $\mathcal{K}$ with the same velocity as that of the external charged
point particle $q_{f}.$ Thus, if considered with respect to the laboratory
reference $\mathcal{K}_{f},$ the external charged particle $q_{f}$ will be
in rest, but the test charged point particle $q$ will be moving with the
resulting velocity $u-u_{f}\in T(\mathbb{E}^{3}),$ where, by definition, $%
u:=dr/dt,$ $u_{f}:=dr_{f}/dt,$ $t\in \mathbb{R},$ are the corresponding
velocities of these charged point particles $q$ and $q_{f}$ with respect to
the laboratory reference system $\mathcal{K}.$ As a result of these
reasonings we can write the following action functional expression%
\begin{equation}
S^{(\tau )}=-\int_{s(\tau _{1})}^{s(\tau _{2})}\bar{W}<\dot{\eta}_{f},\dot{%
\eta}_{f}>_{\mathbb{E}^{4}}^{1/2}ds,  \label{A4.1}
\end{equation}%
where, by definition, $\eta _{f}:=(\tau ,r-r_{f})\in \mathbb{E}^{4}$ is the
charged point particle $q$ position coordinates with respect to the rest
reference system $\mathcal{K}_{r}$ and calculated subject to the introduced
laboratory reference system $\mathcal{K}_{f},$ $s\in \mathbb{R}$
parameterizes the corresponding point particle world line, being
infinitesimally related to the time parameter $t\in \mathbb{R}$ as
\begin{equation}
dt:=<\dot{\eta}_{f},\dot{\eta}_{f}>_{\mathbb{E}^{4}}^{1/2}ds.  \label{A4.2}
\end{equation}%
The boundary conditions for functional (\ref{A4.1}) are taken naturally in
the form%
\begin{equation}
\delta \xi (s(\tau _{1}))=0=\delta \xi (s(\tau _{2}),  \label{A4.3}
\end{equation}%
where $\xi =(\tau ,r)\in \mathbb{E}^{4}.$ The least action condition $\delta
S^{(\tau )}=0$ jointly with (\ref{A4.3}) gives rise to the \bigskip
following equations:%
\begin{eqnarray}
P &:&=\partial \mathcal{L}^{(\tau )}/\partial \dot{\xi}=-\bar{W}\dot{\eta}%
_{f}<\dot{\eta}_{f},\dot{\eta}_{f}>_{\mathbb{E}^{4}}^{-1/2},  \notag \\
dP/ds &:&=\partial \mathcal{L}^{(\tau )}/\partial \xi =-\nabla _{\xi }\bar{W}%
<\dot{\eta}_{f},\dot{\eta}_{f}>_{\mathbb{E}^{4}}^{1/2},  \label{A4.4}
\end{eqnarray}%
where the Lagrangian function equals
\begin{equation}
\mathcal{L}^{(\tau )}:=-\bar{W}<\dot{\eta}_{f},\dot{\eta}_{f}>_{\mathbb{E}%
^{4}}^{1/2}.  \label{A4.5}
\end{equation}%
Having now defined the charged point particle $q$ momentum $p\in T^{\ast }(%
\mathbb{E}^{3})$ as
\begin{equation}
p:=-\bar{W}\dot{r}<\dot{\eta}_{f},\dot{\eta}_{f}>_{\mathbb{E}^{4}}^{-1/2}=-%
\bar{W}u  \label{A4.6}
\end{equation}%
and the induced external magnetic vector potential $A\in T^{\ast }(\mathbb{E}%
^{3})$ as
\begin{equation}
qA:=\bar{W}\dot{r}_{f}<\dot{\eta}_{f},\dot{\eta}_{f}>_{\mathbb{E}%
^{4}}^{-1/2}=\bar{W}u_{f},  \label{A4.7}
\end{equation}%
we obtain, owing to relationship (\ref{A4.2}), the relativistic Lorentz type
force expression%
\begin{equation}
dp/dt=qE+qu\times B-q\nabla <u,A>_{\mathbb{E}^{3}},  \label{A4.8}
\end{equation}%
where we denoted, by definition,
\begin{equation}
E:=-q^{-1}\nabla \bar{W}-\partial A/\partial t,\text{ \ \ \ }B=\nabla \times
A,  \label{A4.9}
\end{equation}%
being, respectively, the external electric and magnetic fields, acting on
the charged point particle $q.$

The result (\ref{A4.8}) contains the additional Lorentz force component
\begin{equation}
F_{c}:=-q\nabla <u,A>_{\mathbb{E}^{3}},  \label{A4.10}
\end{equation}
not present in the classical relativistic Lorentz force expressions (\ref%
{A1.6}) and (\ref{A2.10}), obtained before. Moreover, from (\ref{A4.6}) one
obtains that the point particle $q$ momentum
\begin{equation}
p=-\bar{W}u:=mu,  \label{A4.11}
\end{equation}%
where the particle mass
\begin{equation}
m:=-\bar{W}  \label{A4.12}
\end{equation}%
does not already coincide with the corresponding classical relativistic \
relationship of (\ref{A1.7}).

Consider \ now the least action condition for functional (\ref{A4.1}) at the
critical parameter $s=\tau \in \mathbb{R}:$
\begin{eqnarray}
\delta S^{(\tau )} &=&0,\text{ \ \ \ \ \ }\delta r(\tau _{1})=0=\delta
r(\tau _{2}),  \label{A4.13} \\
S^{(\tau )} &:&=-\int_{\tau _{1}}^{\tau _{2}}\bar{W}(1+|\dot{r}-\dot{r}%
_{f}|^{2})^{1/2}d\tau .  \notag
\end{eqnarray}%
The resulting Lagrangian function
\begin{equation}
\mathcal{L}^{(\tau )}:=-\bar{W}(1+|\dot{r}-\dot{r}_{f}|^{2})^{1/2}
\label{A4.14}
\end{equation}%
gives rise to the generalized momentum expression%
\begin{equation}
P:=\partial \mathcal{L}^{(\tau )}/\partial \dot{r}=-\bar{W}(\dot{r}-\dot{r}%
_{f})(1+|\dot{r}-\dot{r}_{f}|^{2})^{-1/2}:=p+qA,  \label{A4.15}
\end{equation}%
which makes it possible to construct \cite{AM,Ar,PM,DNF,HPP} the
corresponding Hamiltonian function as
\begin{eqnarray}
\mathcal{H} &:&=<P,\dot{r}>_{\mathbb{E}^{3}}-\mathcal{L}^{(\tau )}=-(\bar{W}%
^{2}-|p+qA|^{2})^{1/2}-  \notag \\
- &<&p+qA,qA>_{\mathbb{E}^{3}}(\bar{W}^{2}-|p+qA|^{2})^{-1/2},  \label{A4.16}
\end{eqnarray}%
satisfying the canonical Hamiltonian equations
\begin{equation}
dP/d\tau :=\partial \mathcal{H}/\partial r,\text{ \ \ }dr/d\tau :=-\partial
\mathcal{H}/\partial r,  \label{A4.16a}
\end{equation}%
evolving with respect to the proper rest reference system time $\tau \in
\mathbb{R}$ parameter. When deriving (\ref{A4.16}) we made use of
relationship (\ref{A4.2}) at $s=\tau \in \mathbb{R}$ jointly with
definitions (\ref{A4.6}) and (\ref{A4.7}). Since the Hamiltonian function (%
\ref{A4.16}) is conservative with respect to the evolution parameter $\tau
\in \mathbb{R},$ owing to relationship (\ref{A4.2}) at $s=\tau \in \mathbb{R}
$ one obtains that
\begin{equation}
d\mathcal{H}/d\tau =0=d\mathcal{H}/dt  \label{A4.17}
\end{equation}%
for all $t,\tau \in \mathbb{R}.$ The obtained results one can formulated as
the following proposition.

\begin{proposition}
The charged point particle electrodynamics, related with the least
action principle (\ref{A4.1}) and (\ref{A4.3}), reduces to the
modified Lorentz type force equation (\ref{A4.8}), and is equivalent
to the canonical Hamilton system (\ref{A4.16a}) with respect to the
proper rest reference system time parameter $\tau \in \mathbb{R}.$
The corresponding Hamiltonian
function (\ref{A4.16}) is a conservation law for the Lorentz type dynamics (%
\ref{A4.8}), satisfying the conditions (\ref{A4.17}) with respect to the
both reference systems parameters $t,\tau \in \mathbb{R}.$
\end{proposition}

\bigskip As a corollary, the corresponding energy expression for
electrodynamical model (\ref{A4.8}) can be defined as%
\begin{equation}
\mathcal{E}:=(\bar{W}^{2}-|p+qA|^{2})^{1/2}+<p+qA,qA>_{\mathbb{E}^{3}}(\bar{W%
}^{2}-|p+qA|^{2})^{-1/2}.  \label{A4.18}
\end{equation}%
The obtained above energy expression (\ref{A4.18}) is a necessary ingredient
for quantizing the relativistic electrodynamics (\ref{A4.8}) of our charged
point particle $q$ under the external electromagnetic field influence.

\section{A new hadronic string model: the least action principle and
relativistic electrodynamics analysis within the vacuum field theory approach%
}

\subsection{ A new hadronic string model least action formulation}

A classical relativistic hadronic string model was first proposed in \cite%
{BC,Na,Go} and deeply studied in \cite{BN}, making use of the least action
principle and related Lagrangian and Hamiltonian formalisms. We will not
here discuss this classical string model and not comment physical problems
accompanying it, especially those related with its diverse quantization
versions, but proceed to formulating a new relativistic hadronic string
model, constructed by means of the vacuum field theory approach, devised in
\cite{BPT,BPT1,BP}. The corresponding least action principle is, following
\cite{BN}, formulated as
\begin{equation}
\delta S^{(\tau )}=0,\text{ \ \ }S^{(\tau )}:=\int_{s(\tau _{1})}^{s(\tau
_{2})}ds\int_{\sigma _{1}(s)}^{\sigma _{2}(s)}\bar{W}(x(\xi ))(\dot{\xi}%
^{2}\xi ^{\prime 2}-<\dot{\xi},\xi ^{\prime }>_{\mathbb{E}%
^{4}}^{2})^{1/2}d\sigma \wedge ds,  \label{A5.1}
\end{equation}%
where $\bar{W}:M^{4}\rightarrow \mathbb{R}$ is a vacuum field potential
function, characterizing the interaction of the vacuum medium with our
string object, the differential 2-form $d\Sigma ^{(2)}:=(\dot{\xi}^{2}\xi
^{\prime 2}-<\dot{\xi},\xi ^{\prime }>_{\mathbb{E}^{4}}^{2})^{1/2}d\sigma
\wedge ds$ $=\sqrt[2]{g(\xi )}d\sigma \wedge ds,$ $g(\xi ):=$ $\det (\left.
g_{ij}(\xi )\right\vert _{i,j=\overline{1,2}}),$ $\dot{\xi}^{2}:=$ $<\dot{\xi%
},\dot{\xi}>_{\mathbb{E}^{4}},$ \ $\xi ^{\prime ,2}:=$ $\ <\xi ^{\prime
},\xi ^{\prime }>_{\mathbb{E}^{4}},$ being related with the Euclidean
infinitesimal metrics $dz^{2}:=<d\xi ,d\xi >_{\mathbb{E}^{4}}=g_{11}(\xi
)d\sigma ^{2}+g_{12}(\xi )d\sigma ds+g_{21}(\xi )dsd\sigma $ $+g_{22}(\xi
)ds^{2}$ on the string world surface, means \cite{AM,Th,DNF,BN} the
infinitesimal two-dimensional world surface element, parameterized by
variables $(\sigma ,s)\in \mathbb{E}^{2}$ and embedded into the
4-dimensional Euclidean space-time with coordinates $\xi :=(r,\tau (\sigma
,s)))\in \mathbb{E}^{4}$ subject to the proper rest reference system $%
\mathcal{K},$ $\dot{\xi}:=\partial \xi /\partial s,$ $\xi ^{\prime
}:=\partial \xi /\partial \sigma $ are the corresponding partial
derivatives. The related boundary conditions are chosen as
\begin{equation}
\delta \xi (\sigma (s),s)=0=\delta \xi (\sigma (s),s)  \label{A5.2}
\end{equation}%
for all $s\in \mathbb{R}.$ The action functional expression is strongly
motivated by that constructed for the point particle action functional (\ref%
{A3.1}):
\begin{equation}
S^{(\tau )}:=-\int_{\sigma _{1}}^{\sigma _{2}}dl(\sigma )\int_{t(\sigma
,\tau _{1})}^{t(\sigma ,\tau _{2})}\bar{W}dt(\tau ,\sigma ),  \label{A5.3}
\end{equation}%
where the laboratory reference time parameter $t(\tau ,\sigma )\in \mathbb{R}
$ is related to the proper rest string reference system time parameter $\tau
\in \mathbb{R}$ by means of the standard Euclidean infinitesimal
relationship
\begin{equation}
dt(\tau ,\sigma ):=(1+\dot{r}_{\perp }^{2}(\tau ,\sigma ))^{1/2}d\tau ,
\label{A5.4}
\end{equation}%
with $\sigma \in \lbrack \sigma _{1},\sigma _{2}]$ being a spatial variable
\ parameterizing the string length measure $dl(\sigma )$ on the real axis $%
\mathbb{R},$ $\dot{r}_{\perp }(\tau ,\sigma ):=\hat{N}$ $\dot{r}(\tau
,\sigma )\in \mathbb{E}^{3}$ being the orthogonal to the string velocity
component, and
\begin{equation}
\text{\ }\ \hat{N}:=(1-r^{\prime ,-2}r^{\prime }\otimes r^{\prime }),
\label{A5.5}
\end{equation}%
being the corresponding projector operator in $\mathbb{E}^{3}$ on the
orthogonal to the string direction, expressed for brevity by means of the
standard tensor product $"\otimes "$ in the Euclidean space $\mathbb{E}^{3}.$
The result of calculation of (\ref{A5.3}) gives rise to the following
expression
\begin{equation}
S^{(\tau )}=-\int_{\tau _{1}}^{\tau _{2}}d\tau \int_{\sigma _{1}(\tau
)}^{\sigma _{2}(\tau )}\bar{W}[(r^{\prime ,2}(1+\dot{r}^{2})-<\dot{r}%
,r^{\prime }>_{\mathbb{E}^{3}}]^{1/2}d\ \sigma ,  \label{A5.6}
\end{equation}%
where we made use of the infinitesimal measure representation
$dl(\sigma )=<r^{\prime },r^{\prime }>_{\mathbb{E}^{3}}^{1/2}d\sigma
,$ $\sigma \in \lbrack \sigma _{1},\sigma _{2}].$ If now to
introduce on the string world surface local coordinates $(\sigma
,s(\tau ,\sigma ))\in \mathbb{E}^{2}$ and the related Euclidean
string position vector $\xi :=(r(\sigma ,s),\tau )\in
\mathbb{E}^{4},$ the string action functional reduces equivalently
to that of (\ref{A5.1}).

Below we will proceed to Lagrangian and Hamiltonian analyzing the least
action conditions for expressions (\ref{A5.1}) and (\ref{A5.6}).

\subsection{ Lagrangian and Hamiltonian analysis}

First we will obtain the corresponding to (\ref{A5.1}) Euler
equations with respect to the special \cite{BN,DNF} internal
conformal variables $(\sigma ,s)\in \mathbb{E}^{2}$ on the world
string surface, with respect to which
the metrics on it becomes equal to $dz^{2}=\xi ^{\prime ,2}d\sigma ^{2}+\dot{%
\xi}^{2}ds^{2},$ where $<$ $\xi ^{\prime },\dot{\xi}$ $>_{\mathbb{E}%
^{4}}=0=\xi ^{\prime ,2}-\dot{\xi}^{2},$ and the corresponding infinitesimal
world surface measure $d\Sigma ^{(2)}$ becomes $d\Sigma ^{(2)}=(\xi ^{\prime
,2}\dot{\xi}^{2})^{1/2}d\sigma \wedge ds.$ As a result of simple
calculations one finds the linear second order partial differential equation
\begin{equation}
\partial (\bar{W}\dot{\xi})/\partial s+\partial (\bar{W}\xi ^{\prime
})/\partial \sigma =(\xi ^{\prime ,2}\dot{\xi}^{2})^{1/2}\partial \bar{W}%
/\partial \sigma  \label{A6.1}
\end{equation}%
under the suitably chosen boundary conditions%
\begin{equation}
\xi ^{\prime }-\dot{\xi}\text{ }\dot{\sigma}=0  \label{A6.2}
\end{equation}%
for all $s\in \mathbb{R}.$ It is interesting to mention that the equation \ (%
\ref{A6.1}) \ is of elliptic type, contrary to the case considered before in
\cite{BN}. This is, evidently, owing to the fact that the resulting metrics
on the string world surface is Euclidean, as we took into account that the
real string motion is, in reality, realized with respect to its proper rest
reference system $\mathcal{K}_{r},$ being not dependent on the string motion
observation data, measured with respect to any external laboratory reference
system $\mathcal{K}.$

The differential equation (\ref{A6.1}) strongly depends on the vacuum field
potential function $\bar{W}:M^{4}\rightarrow \mathbb{R},$ which, as a
function of the Minkovski 4-vector variable $x:=(r,t(\sigma ,s))\in M^{4}$
of the laboratory reference system $\mathcal{K},$ should be \ expressed by
means of the infinitesimal relationship (\ref{A5.4}) in the following form:%
\begin{equation}
dt=<\hat{N}\partial \xi /\partial \tau ,\hat{N}\partial \xi /\partial \tau
>^{1/2}(\frac{\partial \tau }{\partial s}ds+\frac{\partial \tau }{\partial
\sigma }d\sigma ),  \label{A6.3}
\end{equation}
defined on the string world surface. The function
$\bar{W}:M^{4}\rightarrow \mathbb{R}$ itself should be
simultaneously found by means of a suitable solution to the Maxwell
equation $\partial ^{2}W/\partial t^{2}-\Delta W=\rho ,$ where $\rho
\in \mathbb{R}$ is an ambient charge density and, by definition, $\
\bar{W}(r(t))$ $:=$ $\lim_{r\rightarrow r(t)}\left.
W(r,t)\right\vert ,$ with $r(t)\in \mathbb{E}^{3}$ being the
position of the string element with coordinates $(\sigma ,\tau )\in
\mathbb{E}^{2}$ at a time moment $t=t(\sigma ,\tau )\in \mathbb{R}.$

Proceed now to constructing the dynamical Euler equations for our string
model, making use of action functional (\ref{A5.6}). It is easy to calculate
that the generalized momentum
\begin{eqnarray}
p &:&=\partial \mathcal{L}^{(\tau )}/\partial \dot{r}=\frac{-\bar{W}%
(r^{\prime 2}\dot{r}-r^{\prime }<r^{\prime },\dot{r}>_{\mathbb{E}^{3}})}{%
[r^{\prime ,2}(\dot{r}^{2}+1)-<r^{\prime },\dot{r}>_{\mathbb{E}%
^{3}}^{2}]^{1/2}}=  \notag \\
&=&\frac{-\bar{W}(r^{\prime 2}\hat{N}\dot{r})}{[r^{\prime ,2}(\dot{r}%
^{2}+1)-<r^{\prime },\dot{r}>_{\mathbb{E}^{3}}^{2}]^{1/2}}  \label{A6.4}
\end{eqnarray}%
satisfies the dynamical equation
\begin{equation}
dp/d\tau :=\delta \mathcal{L}^{(\tau )}/\delta r=[r^{\prime ,2}(\dot{r}%
^{2}+1)-<r^{\prime },\dot{r}>_{\mathbb{E}^{3}}^{2}]^{1/2}\nabla \bar{W}-%
\frac{\partial }{\partial \sigma }\{\bar{W}[(1+\dot{r}^{2}\hat{T})r^{\prime
}]^{-1/2}\},  \label{A6.5}
\end{equation}%
where we denoted by
\begin{equation}
\mathcal{L}^{(\tau )}:=-\bar{W}[(r^{\prime ,2}(1+\dot{r}^{2})-<\dot{r}%
,r^{\prime }>_{\mathbb{E}^{3}}]^{1/2}  \label{A6.6}
\end{equation}%
the corresponding Lagrangian function and by
\begin{equation}
\hat{T}:=1-\dot{r}^{-2}\text{ }\dot{r}\otimes \dot{r}  \label{A6.7}
\end{equation}%
the related dynamic projector operator in $\mathbb{E}^{3}.$ The Lagrangian
function is degenerate \cite{BN,DNF}, satisfying the obvious identity
\begin{equation}
<p,r^{\prime }>_{\mathbb{E}^{3}}=0  \label{A6.8}
\end{equation}%
for all $\tau \in \mathbb{R}.$ Concerning the Hamiltonian formulation of the
dynamics (\ref{A6.5}) we construct the corresponding Hamiltonian functional
as
\begin{eqnarray}
\mathcal{H} &:&=\int_{\sigma _{1}}^{\sigma _{2}}(<p,\dot{r}>_{\mathbb{E}%
^{3}}-\mathcal{L}^{(\tau )})d\sigma =  \notag \\
&=&\int_{\sigma _{1}}^{\sigma _{2}}\{\bar{W}r^{\prime ,2}[r^{\prime ,2}(\dot{%
r}^{2}+1)-<r^{\prime },\dot{r}>_{\mathbb{E}^{3}}^{2}]^{-1/2}d\sigma =  \notag
\\
&=&\int_{\sigma _{1}}^{\sigma _{2}}[(\bar{W}r^{\prime
})^{2}-p^{2}]^{1/2}d\sigma ,  \label{A6.9}
\end{eqnarray}%
satisfying the canonical equations
\begin{equation}
dr/d\tau :=\delta \mathcal{H}/\delta p,\text{ \ }dp/d\tau :=-\delta \mathcal{%
H}/\delta r,  \label{A6.10}
\end{equation}%
where
\begin{equation}
d\mathcal{H}/d\tau =0,  \label{A6.10a}
\end{equation}%
holding only with respect to the proper rest reference system \ $\mathcal{K}%
_{r}$ time parameter $\tau \in \mathbb{R}.$ Making now use of identity (\ref%
{A6.8}) the Hamiltonian functional (\ref{A6.9}) can be equivalently
represented as
\begin{equation}
\mathcal{H}=\int_{\sigma _{1}}^{\sigma _{2}}|\bar{W}r^{\prime }-p|d\sigma .
\label{A6.11}
\end{equation}%
Moreover, concerning the obtained above result we need to mention here that
one can not construct the suitable Hamiltonian function expression and
relationship of type (\ref{A6.10a}) with respect to the laboratory reference
system $\mathcal{K},$ since the expression (\ref{A6.11}) is not defined on
the whole for a separate laboratory time parameter $t\in \mathbb{R}$ locally
dependent both on the spatial parameter $\sigma \in \mathbb{R}$ and the
proper rest reference system time parameter $\tau \in \mathbb{R}.$

Thereby, one can formulate the following proposition.

\begin{proposition}
The hadronic string model (\ref{A5.1})  allows  on the related world
surface the conformal local coordinates, with respect to which the
resulting dynamics is described by means of the linear second order
elliptic equation (\ref{A6.1}). Subject to the proper rest reference
system Euclidean coordinates the corresponding dynamics is
equivalent to the canonical
Hamiltonian equations (\ref{A6.10}) with   Hamiltonian functional  (\ref%
{A6.9}).
\end{proposition}

Proceed now to constructing the action functional expression for a charged
string under an external magnetic field, generated by a point velocity
charged particle $q_{f},$ moving with some velocity $u_{f}:=dr_{f}/dt\in $ $%
\mathbb{E}^{3}$\ subject to a laboratory reference system $\mathcal{K}.$ To
solve this problem we make use of the trick of Section 2 above, passing to
the proper rest reference system $\mathcal{K}_{r}$ with respect to the
relative reference system $\mathcal{K}_{f},$ moving with velocity $u_{f}\in $
$\mathbb{E}^{3}.$ As a result of this reasoning we can write down the action
functional:
\begin{equation}
S^{(\tau )}=-\int_{\tau _{1}}^{\tau _{2}}d\tau \int_{\sigma _{1}(\tau
)}^{\sigma _{2}(\tau )}\bar{W}[r^{\prime ,2}(1+(\dot{r}-\dot{r}_{f})^{2})-<%
\dot{r}-\dot{r}_{f},r^{\prime }>_{\mathbb{E}^{3}}^{2}]^{1/2}d\ \sigma ,
\label{A6.12}
\end{equation}%
giving rise to the following dynamical equation
\begin{eqnarray}
dP/d\tau  &:&=\delta \mathcal{L}^{(\tau )}/\delta r=-[r^{\prime ,2}(1+(\dot{r%
}-\dot{r}_{f})^{2})-<\dot{r}-\dot{r}_{f},r^{\prime }>_{\mathbb{E}%
^{3}}^{2}]^{1/2}\nabla \bar{W}+  \label{A6.13} \\
&&+\frac{\partial }{\partial \sigma }\left\{ \frac{\bar{W}[1+(\dot{r}-\dot{r}%
_{f})^{2}\hat{T}_{f}]r^{\prime }}{[r^{\prime ,2}(1+(\dot{r}-\dot{r}%
_{f})^{2})-<\dot{r}-\dot{r}_{f},r^{\prime }>_{\mathbb{E}^{3}}^{2}]^{1/2}}%
\right\} ,  \notag
\end{eqnarray}%
where the generalized momentum
\begin{equation}
P:=\frac{-\bar{W}[r^{\prime 2}\hat{N}(\dot{r}-\dot{r}_{f})]}{[r^{\prime
,2}(1+(\dot{r}-\dot{r}_{f})^{2})-<\dot{r}-\dot{r}_{f},r^{\prime }>_{\mathbb{E%
}^{3}}^{2}]^{1/2}}  \label{A6.14}
\end{equation}%
and the projection operator in $\mathbb{E}^{3}$
\begin{equation}
\hat{T}_{f}:=1-(\dot{r}-\dot{r}_{f})^{-2}\text{ }(\dot{r}-\dot{r}%
_{f})\otimes (\dot{r}-\dot{r}_{f}).  \label{A6.15}
\end{equation}%
Having defined by
\begin{equation}
p:=\frac{-\bar{W}(r^{\prime 2}\hat{N}\dot{r})}{[r^{\prime ,2}(1+(\dot{r}-%
\dot{r}_{f})^{2})-<\dot{r}-\dot{r}_{f},r^{\prime }>_{\mathbb{E}%
^{3}}^{2}]^{1/2}}  \label{A6.16}
\end{equation}%
the local string momentum and by
\begin{equation}
qA:=\frac{\bar{W}(r^{\prime 2}\hat{N}\dot{r}_{f})}{[r^{\prime ,2}(1+(\dot{r}-%
\dot{r}_{f})^{2})-<\dot{r}-\dot{r}_{f},r^{\prime }>_{\mathbb{E}%
^{3}}^{2}]^{1/2}}  \label{A6.17}
\end{equation}%
the external vector magnetic potential, equation (\ref{A6.13}) reduces to%
\begin{equation}
\begin{array}{c}
dp/d\tau =q\dot{r}\times B-q\nabla <A,\dot{r}>_{\mathbb{E}^{3}}-q\frac{%
\partial A}{\partial \tau }- \\
-[r^{\prime ,2}(1+(\dot{r}-\dot{r}_{f})^{2})-<\dot{r}-\dot{r}_{f},r^{\prime
}>_{\mathbb{E}^{3}}^{2}]^{1/2}\nabla \bar{W}+ \\
+\frac{\partial }{\partial \sigma }\left\{ \frac{\bar{W}[1+(\dot{r}-\dot{r}%
_{f})^{2}\hat{T}_{f}]r^{\prime }}{[r^{\prime ,2}(1+(\dot{r}-\dot{r}%
_{f})^{2})-<\dot{r}-\dot{r}_{f},r^{\prime }>_{\mathbb{E}^{3}}^{2}]^{1/2}}%
\right\} ,%
\end{array}
\label{A6.18}
\end{equation}%
where $q\in \mathbb{R}$ is a charge density, distributed along the string
length, $B:=\nabla \times A$ means the external magnetic field, acting on
the string. The expression, defined as
\begin{eqnarray}
E &:&=-q\frac{\partial A}{\partial \tau }-[r^{\prime ,2}(1+(\dot{r}-\dot{r}%
_{f})^{2})-<\dot{r}-\dot{r}_{f},r^{\prime }>_{\mathbb{E}^{3}}^{2}]^{1/2}%
\nabla \bar{W}+  \label{A6.19} \\
&&+\frac{\partial }{\partial \sigma }\left\{ \frac{\bar{W}[1+(\dot{r}-\dot{r}%
_{f})^{2}\hat{T}_{f}]r^{\prime }}{[r^{\prime ,2}(1+(\dot{r}-\dot{r}%
_{f})^{2})-<\dot{r}-\dot{r}_{f},r^{\prime }>_{\mathbb{E}^{3}}^{2}]^{1/2}}%
\right\} ,  \notag
\end{eqnarray}%
similarly to the charged point particle case, models a related electric
field, exerted on the string by the external electric charge $q_{f}.$ Making
use of the standard scheme, one can derive, as above, the Hamiltonian
interpretation of dynamical equations \ (\ref{A6.13}), on which we will not
stay here in more details.

\section{Conclusion}

Based on the vacuum field theory approach, devised recently in \cite%
{BPT,BPT1,BP}, we revisited the alternative charged point particle and
hadronic string electrodynamics models, having succeeded in treating their
Lagrangian and Hamiltonian properties. The obtained results were compared
with classical ones, owing to which a physically motivated choice of a true
model was argued. Another important aspect of the developed vacuum field
theory approach consists in singling out the decisive role of the related
rest \ reference system $\mathcal{K}_{r},$ with respect to which the
relativistic object motion, in reality, realizes. Namely, with respect to
the proper rest reference \ system evolution parameter $\tau \in \mathbb{R}$
all of our electrodynamics models allow both the Lagrangian and Hamiltonian
physically reasonable formulations, suitable for the canonical procedure.
The deeper physical nature of this fact remains, by now, we assume, not
enough understood. We would like here to recall  only a very interesting
reasonings by R. Feynman who argued in \cite{Fe1,Fe2} that the relativistic
expressions have physical sense only with respect to \ the proper rest
reference systems. In a sequel of our work we plan to analyze our
relativistic electrodynamic models subject to their quantization and make a
step toward the related vacuum quantum field theory of infinite many
particle systems.

\section{Acknowledgments}

Authors are cordially thankful to the Abdus Salam International Centre for
Theoretical Physics in Trieste, Italy, for the hospitality during their
research 2007-2008 scholarships. A.P. is, especially, grateful to Profs. D.
L. Blackmore (NJIT, USA) and P.I. Holod (Kyiv, UKMA) for fruitful
discussions, useful comments and remarks. Last but not least thanks go to
Prof. A.A. Logunov (Moscow, INP) for his interest to the work, as well to
Mrs. Dilys Grilli (Trieste, Publications office, ICTP) and Natalia K.
Prykarpatska for professional help in preparing the manuscript for
publication.

\end{document}